\documentclass[12pt]{iopart}

\begin{document}
\title{Correlators of $N=1$ Superconformal Currents}

\author{Steven H. Simon
}                     
\address{Alcatel-Lucent Bell Labs, Murray Hill NJ 07974, USA}
\begin{abstract}
We give an explicit expression for the $M$-point correlator of the superconformal current in two dimensional $N=1$ superconformal field theories.
\end{abstract}

\maketitle

\section{Main Result}
$N=1$ superconformal field theories\cite{YellowBook} have a chiral algebra generated by a spin 2 field $T$ known as the stress energy tensor and a spin 3/2 field $G$ known as the superconformal current which is the superpartner of $T$.  The singular terms of the operator product expansion of these fields are given by
\begin{eqnarray}
    T(z) T(z') &=&  \frac{c/2}{(z - z')^4} + \frac{2 T(z')}{(z -z')^2} + \frac{\partial T(z')}{z - z'}  + {\cal O}(1) \\
    \label{eq:2} T(z) G(z') &=&  \frac{(3/2)G(z') }{(z-z')^2} \,   +  \frac{\partial G(z')}{z-z'}    + {\cal O}(1) \\
 \label{eq:3}   G(z) G(z') &=&   \frac{(2 c/3) }{(z-z')^3} \,   +  \frac{2 T(z')}{z-z'}   + {\cal O}(1)
\end{eqnarray}
where $c$ is the central charge.  The form written here is generic so long as there are no further conserved currents\cite{ZamW3}.

The point of this paper is to give an expression for the correlator
\begin{equation}
    C_M = \langle G(z_1) G(z_2) \ldots  G(z_M) \rangle
\end{equation}
where $M$ must be even or the correlator will vanish.   Obviously the two point correlator is directly given by the OPE as
\begin{equation}
\label{eq:start}
C_2  =   \frac{(2 c/3) }{(z_1-z_2)^3}
\end{equation}

Using Eq. \ref{eq:2} and \ref{eq:3} we can examine the pole structure with respect to one variable $z_1$, and derive a recursion relation for this correlator  (compare for example with the discussions in Ref.~\cite{ZamW3} and \cite{WenWu}) whereby the $M$ point correlator is written in terms of the $M-2$ point correlator
\begin{eqnarray}
\label{eq:recursion}
    && C_M = \\ && \sum_{i=2}^M   (-1)^i {\mbox{\Huge$($}} \underbrace{\rule[-15pt]{0pt}{15pt} \frac{ (2 c/3)}{(z_1 - z_i)^3}}_{\alpha}  + \frac{2}{(z_1 - z_i)} \sum_{j \neq 1,i}  {\mbox{\Huge$[$}} \underbrace{\rule[-15pt]{0pt}{15pt}  \frac{3/2}{(z_i - z_j)^2} }_{\beta}+ \underbrace{\rule[-15pt]{0pt}{15pt}\frac{1}{z_i - z_j} \frac{\partial}{\partial {z_j}}}_\gamma {\mbox{\Huge$]$}}  {\mbox{\Huge{$)$}}}
      C_{M-2}(\hat 1, \hat i) \nonumber
\end{eqnarray}
where $C_{M-2}(\hat 1, \hat i)$ is the $M-2$ point correlator that excludes coordinates $z_1$ and $z_i$
\begin{equation}
  C_{M-2}(\hat 1, \hat i) = \langle G(z_2)  \ldots  G(z_{i-1}) G(z_{i+1}) \ldots G(z_M) \rangle
\end{equation}
The terms in Eq.~\ref{eq:recursion} are labeled $\alpha, \beta, \gamma$ for future reference.  Thus through this recursion relation (and starting the recursion with Eq.~\ref{eq:start}) in principle the $M$ point correlator can be determined.

We claim that the solution to this recursion relation can be written in a very simple closed form.   We write the correlator in terms of a fully analytic ``wavefunction\cite{MooreRead}" $\psi_M$
\begin{equation}
    C_M = \psi_M \,\, V_M^{-3}
\end{equation}
where
\begin{equation}
    V_M = \prod_{1 \leq i < j \leq M} (z_i -z_j)
\end{equation}
Inspired by the construction of\cite{ReadRezayi}, we write the wavefunction as
\begin{equation}
    \psi_M = {\cal N} \sum_{P \in S_M} \,\, \prod_{1 \leq r < s \leq M/2}  \chi(z_{P(2r-1)}, z_{P(2r)};  z_{P(2s-1)}, z_{P(2s)})
\end{equation}
with the normalization constant
\begin{equation}
{\cal N} = \frac{(c/3)^{(M/4)(3-M/2)}}{(M/2)!}
\end{equation}
Here, $P$ is a permutation of the integers $1 \ldots M$ and the sum is over all such permutations.   The function $\chi$ is given by
\begin{eqnarray} \nonumber
 \chi(z_1, z_2; z_3, z_4) &=& A (z_1 - z_3)^3 (z_2 - z_4)^3 (z_1 - z_4)^3 (z_2 - z_3)^3
\label{eq:N1chi} \\  & &
+ (z_1 - z_3)^4 (z_2 - z_4)^4 (z_1 - z_4)^2 (z_2 - z_3)^2
\end{eqnarray}
where    $A = (c/3) - 1$.  Perhaps a more useful form is given by merging $V$ with $\psi$ to write directly
\begin{eqnarray}
\nonumber  C_M =  \frac{1}{(M/2)!} \sum_{P \in S_M} \sigma(P) & & \left[ \! \prod_{1 \leq r < s \leq M/2}  \tilde \chi(z_{P(2r-1)}, z_{P(2r)};  z_{P(2s-1)}, z_{P(2s)}) \right.  \\ & & ~~~~ \left. \prod_{1 \leq t \leq M/2} \frac{(c/3)}{(z_{P(2t -1)} - z_{P(2t)})^{3}} \right] \label{eq:secondform}
\end{eqnarray}
where $\sigma(P)$ is the signature $(\pm1)$ of the permutation and
 where $\tilde \chi$ is given by
\begin{eqnarray}
\nonumber
\tilde \chi(z_1, z_2; z_3, z_4) &=& \frac{3}{c} \left[ A  + \frac{(z_1 - z_3) (z_2 - z_4)}{ (z_1 - z_4) (z_2 - z_3)}
 \right]         \\ &=&
      \rule[0pt]{0pt}{20pt}  1 +  (3/c) \frac{(z_1 - z_2) (z_3 - z_4)}{ (z_1 - z_4) (z_2 - z_3)}      \label{eq:tildechi}
\end{eqnarray}
Using Mathematica\texttrademark, we have been able to verify that this form of the correlator satisfies the recursion relation for $M \leq 10$.  Below we sketch a proof that this recursion relation holds for general $M$.

We note in passing that correlators of spin 1 and spin 2 fields of chiral algebras have been found to have somewhat similar (but in some ways even simpler) forms.  See the discussions in Ref.~\cite{Gaberdiel} of the work of Ref.~\cite{Frenkel}.

\section{Special Cases and Further Checks}
There are several special cases where $C_M$ may be calculated or tested by other means to check the validity of our result.   These cases are $c=1$, $3/2$, $-6/5$, $7/10$, $-21/4$, and $\infty$.

{\it Free Boson:} The $c=1$ case is a free boson theory where $G$ is given by
\begin{equation}
    G \sim  [ : e^{i \sqrt{3} \phi(z)} :  - : e^{-i \sqrt{3} \phi(z)}: ]
\end{equation}
where $\phi$ is a free boson (See for example Ref.~\cite{Kiritsis}).   To evaluate the correlator of many $G$ fields, we multiply out the terms in all combinations.  The only terms that can be nonzero, must have overall charge neutrality\cite{YellowBook} so that the number of terms $e^{i \sqrt{3} \phi}$ must be the same as the number of terms of $e^{-i \sqrt{3} \phi}$.     Thus all terms are of the form
\begin{equation}
    \prod_{i,j \in A ; i<j} (z_i - z_j)^{3} \prod_{i,j \in B; i<j} (z_i - z_j)^3 \prod_{i \in A, j \in B} (z_i - z_j)^{-3}
\end{equation}
where all of the particles are divided up into two groups $A$ and $B$ of equal size (compare this form to that used in Ref.~\cite{Cappelli}).  Summing all such terms should be equivalent to our predicted form of $C_M$ for $c=1$.  This has been verified numerically for $M\leq 12$.

{\it Ising Cubed:} The $c=3/2$ case can be thought of as the cube of the Ising CFT, the Virasoro minimal model $M(3,4)$, which has $c=1/2$.   $G$ is simply the product of the three Majorana fields, one from each theory.  The correlator of Majorana fields in $M(3,4)$ is a simple Pfaffian\cite{YellowBook} written as ${\rm Pf}[1/(z_i - z_j)]$ thus the correlator of the $G$ fields should be the cube of this Pfaffian.   Again we can verify numerically for $M \leq 12$ particles that our predicted form for $C_M$ is indeed the cube of this Pfaffian for $c=3/2$.

{\it $M(3,5)$ Squared:}
\label{sub:gaffsquarecheck}
The case $c=-6/5$ can be thought of as the square of the Virasoro minimal model $M(3,5)$ which has $c=-3/5$.  Each $M(3,5)$ theory contains a field with spin 3/4, and the field $G$ is then the product of these two fields.   The correlator of the fields with weight 3/4 in the $M(3,5)$ theory has been given explicitly in Ref.~\cite{Gaffnian}, and the correlator of the $G$ fields is then the square of this quantity.   Again we verify numerically for $M \leq 12$ particles that our predicted form for $C_M$ with $c =-6/5$ agrees with this quantity.

{\it Tricritical Ising:}
The case of $c=7/10$ is the tricritical Ising\cite{YellowBook} model, the Virasoro minimal model $M(4,5)$ where $G$ corresponds to the field $\phi_{3,1}$.  The four-point correlator of this field has been calculated in Ref.~\cite{Mattis} and indeed agrees with our expression.  Generally, the many-point correlator of this field must satisfy a particular differential equation corresponding to a null vector condition (See Eq.~D8 from BPZ\cite{BPZ}).   One can check that our proposed form for $C_M$ with $c=7/10$ does indeed satisfy this differential equation for small $M$.  This has been verified for $M\leq 8$ particles using Mathematica\texttrademark.

{\it $M(3,8)$}:   The Virasoro minimal model $M(3,8)$ is the $c=-21/4$ case where $G$ corresponds to the field $\phi_{2,1}$.  This field must similarly satisfy a differential equation corresponding to a null vector condition (Eq. 5.17 of BPZ\cite{BPZ}).  One can again check that our proposed form for $C_M$ satisfies this condition for $M \leq 8$ using Mathematica\texttrademark.

{\it ``Semiclassical" Limit}: The limit of $c=\infty$ is trivial to solve.  In this limit, our proposed expression for the correlator, Eq.~\ref{eq:secondform}, takes the form of a Pfaffian ${\rm Pf}[1/(z_i - z_j)^3]$.   In the $c=\infty$ limit only the $\alpha$ term survives in the recursion relation Eq.~\ref{eq:recursion}.  It is then easy to see that this proposed Pfaffian form does indeed satisfy this simplified recursion relation.

\section{Sketch of Proof}
In this section we prove that the recursion relation Eq.~\ref{eq:recursion} holds for general $M$ (for small $M$ it is easy to check explicitly).  The proof is messy but shows how, with sufficient algebra, the recursion relation can be established.  It seems likely a more elegant proof may also be found.

To check the recursion relation in general, we examine the poles that occur when $z_1$ approaches some other particle $z_i$ as in Eq.~\ref{eq:recursion}.   For simplicity we consider the case of $i=2$.

Examining  the form of Eq.~\ref{eq:secondform} and Eq.~\ref{eq:tildechi}, it is clear that the only way to obtain a third order pole when 2 approaches 1 is when the permutation pairs particles 1 and 2  so that $P(2 r)=1$ and $P(2r-1) =2$ or vice versa.   Further, from Eq.~\ref{eq:tildechi} it is clear that in the product of $\tilde \chi(1,2; m, n)$ terms only the ``$1$" term in Eq.~\ref{eq:tildechi} can contribute from $\tilde \chi$ or the pole will be lower order.  Counting the number of permutations that contribute, we see that the coefficient of the third order pole is  $C_{M-2}(\hat 1,\hat 2)$ with precisely the right coefficients to account for  the $\alpha$ term of Eq.~\ref{eq:recursion}.

We next must show that there is no second order pole as $z_2$ approaches $z_1$ (since there is no second order pole in Eq.~\ref{eq:recursion}).  Again, using the form of Eq.~\ref{eq:secondform} and Eq.~\ref{eq:tildechi} it is clear that the only way to obtain a second order pole is again when the permutation pairs particles 1 and 2  so that $P(2 r)=1$ and $P(2r-1) =2$ or vice versa, and then when we multiply out the product of the $\tilde \chi$ terms we should include exactly one factor of $(z_1 - z_2)(z_m -z_n)/(z_1 - z_n)(z_2 - z_m)$.  Keeping such terms then taking the limit of $z_1 \rightarrow z_2$ we see that the resulting expression becomes symmetric under interchange of $1$ and $2$ to leading order.  However the sum in Eq.~\ref{eq:secondform} is antisymmetrized because of the factor of $\sigma(P)$ thus canceling this term.

Finally we turn to the single pole as $z_1$ approaches $z_2$.  There are three ways to get a first order pole.   In the first two ways, 1 and 2 are in the same pair, and terms in the product of $\tilde \chi$'s give $(z_1 - z_2)^2$ which then combines with the $1/(z_1 - z_2)^3$ term in Eq.~\ref{eq:secondform} to give a single pole.   One way to get such a quadratic term is by expanding terms of the form (here we use shorthand notation  $(12) = z_1 - z_2$)
\begin{equation}
    \label{eq:t1}
    \lim_{1 \rightarrow 2} \tilde \chi(1,2; 3,4) =  1 + (3/c) \frac{(12)(34)}{(24)(23)} - (3/c) \frac{(12)^2(34)}{(24)^2(23)} + \ldots
\end{equation}
to second order as shown.  A second way is via the product of two $\tilde \chi$'s
\begin{equation}
\label{eq:t2}
   \lim_{1 \rightarrow 2} \tilde \chi(1,2; 3,4) \tilde \chi(1,2; 5,6) = \ldots + (3/c)^2 \frac{(12)^2 (34)(56)}{(23)(24)(25)(26)}   +  \ldots
\end{equation}
The third way to get a first order pole is when particles 1 and 2 are not paired, in which case we have terms of the type
\begin{equation}
\label{eq:t3}
   \lim_{1 \rightarrow 2} \tilde \chi(1,4; 3,2)  = 1 + (3/c) \frac{(24)(32)}{(12)(43)} + \ldots
\end{equation}
Note that in all three equations we have replaced 1 with 2 in all places (since we want the evaluation at the pole) except to make the (12) factors explicit.   The sum of the coefficients of these residues should give the $\beta$ and $\gamma$ terms of the recursion relation Eq.~\ref{eq:recursion}.

Once we have isolated the residues of these poles, we would like to find the further poles with respect to the approach of a third particle towards the second (the third particle is $j$ in Eq.~\ref{eq:recursion}).   In Eq.~\ref{eq:t1} there is a second order pole for $j=4$.  Then taking position 4 to approach 2, the coefficient of this pole is then just the product of all $\tilde \chi$ that are not a function of positions 1 or 2 (we use only the ``$1$" term from any other $\tilde \chi(1,2; n, m)$ so we do not pick up any more powers of (12)) times the factors of $1/(z-z)^3$ not including $1/(12)^3$.  Summing over these obviously gives $C_{M-2}(\hat 1, \hat 2)$ however the coefficient is only 1/3 that necessary to account for the coefficient of term $\beta$.    Let us call this contribution term A.

To evaluate the first order pole as 2 approaches 4, we expand the numerator of Eq.~\ref{eq:t1} so that $(34)= (32)+(24)$.  Having first taken the (12) residue, and the first order pole as 4 approaches 2 then gives $-1/(43)$ times all terms not including $1$ and $2$.     There is obviously another way to get a first order pole from Eq.~\ref{eq:t1} by taking $j=3$  (ie., when 3 approaches 2), in this case the residue is exactly the same as the previous case.  Let us call these contributions terms B.

We now turn to the poles that arise from the form of Eq.~\ref{eq:t2}.  Here, there is no possibility of getting  a second order pole as any particle (excluding 1) approaches $2$.  The single poles are for example, when 2 approaches 4 we have a residue $-(56)/(45)(46)$ times factors of $\tilde \chi$ that do not contain 1 and 2 and times the $1/(z-z)^3 $ factors not including $1/(12)^3$.   Call these terms C.   Note that as compared to term $B$, these terms contain one additional power of $3/c$.

We now examine terms of the type of Eq.~\ref{eq:t3} which are the most complicated.  As written there, the relevant permutation has $P(1)=1, P(2)=4, P(3)=3, P(4)=2$ and let us assume $P(i) = i$ for $i > 4$ for simplicity here.   The relevant term of the permutation sum is of the form
\begin{eqnarray} \nonumber
& & \frac{(c/3)^2 \tilde \chi(1,4,3,2)}{(z_1 -z_4)^3 (z_3 - z_2)^3} \left[ \prod_{n > 2} \tilde \chi(1,4; 2n-1, 2n) \right] \left[ \prod_{n > 2} \tilde \chi(3,2; 2n-1, 2n) \right] \times \\ & & \left[\prod_{n > m > 2} \tilde \chi(2m-1, 2m; 2n-1, 2n) \right]\left[ \prod_{n > 2} (c/3)(z_{2n-1} - z_{2n})^{-3} \right] \label{eq:three}
\end{eqnarray}
As 2 approaches 1 there is a single pole, as expressed in Eq.~\ref{eq:t3}.   Other than this pole, any occurance of 1 in this equation is then replaced by 2  (since we are looking at the residue of the pole).  Then due to the factor of $1/(24)^3$ out front there will be a second order pole as 4 approaches 2 with a prefactor of $-(c/3)/(43)(32)^2$ times the terms in brackets.  Considering the first term in brackets the 1 is replaced by 2, and in multiplying out these factors of $\tilde \chi$ we must only keep the ``$1$" term since the other terms have a factor of (24) and reduce the order of the (24) pole.  The coefficient of the second order pole as 2 approaches 4 then is given by the remaining factors with 2 replaced by 4 everywhere it occurs.  This then results in precisely the same quantity (and same sign once we include the $\sigma(P)$ factor) as in term A above.  Further there will be an identical term from exchanging coordinates 3 and 4, leaving 1 and 2 alone, and exchanging each additional coordinate with its partner  ($P(2n) \leftrightarrow P(2n-1)$).  Adding these two terms with term A above gives precisely the right coefficient to account for term $\beta$ in Eq.~\ref{eq:recursion}.

In terms of the type of Eq.~\ref{eq:t3} considering the residue evaluated when 1 approaches 2, there is also the possibility of a first order pole as particle 2 approaches another coordinate.  There are two ways this may occur.   The first way is similar to the above paragraph, only here we take a subleading term, expanding 2 around the position of 4.   (i.e, we consider the second order pole $1/(24)^2$ as in the previous paragraph, but we expand coordinate 2 around position 4 so that at first order there is also a factor of $(24)$ in the numerator).  The coefficient of the first order pole of 2 with respect to 4 is then
\begin{equation} \label{eq:der}
 \left.\frac{\partial}{\partial z_2}\right|_{z_2 \rightarrow z_4} \left\{  \frac{(32)}{(43)}   \frac{(c/3)}{(32)^3}  \left[ \prod_{n > 2} \tilde \chi(2,4; 2n-1, 2n) \right] \left[ \,\, \rule[0pt]{0pt}{15pt}\right]\,\, \left[ \,\, \rule[0pt]{0pt}{15pt}\right] \,\,\left[ \,\, \rule[0pt]{0pt}{15pt}\right] \right\}
 \end{equation}
where the three empty brackets are the same as the last three brackets in Eq.~\ref{eq:three}.  Let us see first what happens when the derivative acts on the outside factor of (32).  The derivative is a trivial $-1$, and we simply replace 2 by 4 everywhere it occurs in the remaining expression.  Note that in the first bracketed expression, when $ 2 \rightarrow 4$ only the ``$1$" term survives. The remaining expression looks just like the terms that would contribute to $C_{M-2}(\hat 1,\hat 2)$ except for the $1/(43)$ out front.  We identify these terms as being exactly the B terms discussed above with a sign that precisely cancels them.  Similarly, let us examine what happens when the derivative acts on the first bracketed quantity.  Here, we get terms of the form $(2n-1,2n)/(4,2n-1)(4,2n)$ times something that looks exactly like our previous expression without the $1/(43)$ and with one additional factor of $3/c$.  We similarly identify these terms as being exactly the C terms discussed above with a sign to cancel them.   We now let the derivative act on  the remaining terms in Eq.~\ref{eq:der}.  Once we take 2 to 4, and sum over all terms we realize that this is precisely $\partial_{z_4} C_{M-2}(\hat 1,\hat 2)$ which correctly gives us the final $\gamma$ term of Eq.~\ref{eq:recursion}.

To complete the proof, we realize (as mentioned above) that there is one more possible source of first order poles as particle 2 approaches other particles in terms of the type of Eq.~\ref{eq:t3} (Here we have already taken the residue of the (12) pole and we are are considering 2 approaching particles other than 1 or 4).   Here we note that such terms all sum to zero.   The terms in question are of the type $\tilde \chi(1,4; 3,2) \tilde \chi(3,2; 5, 6) \tilde \chi(1,4; 5,6)$.  From the first term there is a single pole as 2 approaches 1, then from the second term there is then a single pole as, say, 5 approaches 2.   However, it is easy to show that this combination will be precisely canceled by the same poles that occur in the term of the form  $\tilde \chi(1,6; 3, 2) \tilde \chi(3,2,5,4) \tilde \chi(1,6,5,4)$ (with all remaining coordinates unchanged).

\ack The author is indebted to N. Read and P. Fendley both for helpful conversations and for encouraging him to publish this work.

\section*{References}

%

\end{document}